\documentclass[conference]{IEEEtran}
\IEEEoverridecommandlockouts
\usepackage{color}
\usepackage{graphicx} 
\usepackage{psfrag}  
\usepackage{algorithm}
\usepackage{algorithmic}
\usepackage{amsfonts,amssymb}
\usepackage{amsthm}
\usepackage{amstext}
\usepackage{bm}
\usepackage{subfigure}
\usepackage{hyperref}
\usepackage{amsmath}
\usepackage{amssymb}
\usepackage{siunitx}
\usepackage{array}
\usepackage{booktabs}
\usepackage{amsmath}
\usepackage{algorithm}
\usepackage{algorithmic}
\usepackage{multirow}

\usepackage{epstopdf}

\begin{document}
	
	\title{ QoS-Aware 3D Coverage Deployment of UAVs for Internet of Vehicles in Intelligent Transportation }
	\author{Pengfei~Du,
		Tingyue~Xiao,
		Haotong~Cao,
		Daosen~Zhai
		\thanks{ This work was supported in part by the National Natural Science Foundation of China under Grant 62301447. This work was also supported in part by  the Natural Science Foundation of Sichuan Province under Grant 2023NSFSC1377.  This work was also supported by the Researchers Supporting Project Number (RSPD2023R800), King Saud University, Riyadh, Saudi Arabia}
		\thanks{P. Du and T. Xiao are with the Engineering Research	Center of Intelligent Airground Integrated Vehicle and Traffic Control, Ministry of Education, Xihua University, Chengdu 610039, China. (Email: dupf07010113@163.com and xiaoty1219@163.com).}
		\thanks{H. Cao is with the College of Telecommunications and Information Engineering, Nanjing University of Posts and Telecommunications, Nanjing 210003, China. (Email: haotong.cao1992@gmail.com).}
		\thanks{D. Zhai is with the School of Electronics and Information, Northwestern Polytechnical University, Xi’an 710072, China. (Email: zhaidaosen@nwpu.edu.cn).}
	}
	
	\maketitle
	\begin{abstract}
		It is a challenging problem to characterize the air-to-ground (A2G) channel and identify the best deployment location for 3D UAVs with the QoS awareness.  To address this problem,  we propose a QoS-aware UAV 3D coverage deployment algorithm,  which simulates the three-dimensional urban road scenario, considers the UAV communication resource capacity and vehicle communication QoS requirements comprehensively, and then  obtains the optimal UAV deployment position by improving the genetic algorithm.  Specifically,  the K-means clustering algorithm is used to cluster the vehicles, and the center locations of these clusters serve as the initial UAV positions to generate the initial population. Subsequently, we employ the K-means initialized grey wolf optimization (KIGWO) algorithm to achieve the UAV location with an optimal fitness value by performing an optimal search within the grey wolf population.  To enhance the algorithm's diversity and global search capability, we randomly substitute this optimal location with one of the individual locations from the initial population. The fitness value is determined by the total number of vehicles covered by UAVs in the system, while the allocation scheme's feasibility is evaluated based on the corresponding QoS requirements. Competitive selection operations are conducted to retain individuals with higher fitness values, while crossover and mutation operations are employed to maintain the diversity of solutions. Finally, the individual with the highest fitness, which represents the UAV deployment position that covers the maximum number of vehicles in the entire system, is selected as the optimal solution. Extensive experimental results demonstrate that the proposed algorithm can effectively  enhance the reliability and  vehicle communication QoS.  
	\end{abstract}
	
	
	\section{Introduction}
	The internet of vehicles (IoV) represents a dynamic network that has emerged as an evolution from vehicular ad-hoc networks (VANET).  Within the IoV, vehicles communicate with each other utilizing an extensive array of both short-range and long-range wireless technologies \cite{ref1}. The authors in \cite{ref2} have categorized the communication model into two distinct classifications: vehicle-to-vehicle communication (V2V) and vehicle-to-infrastructure communication (V2I). Nonetheless, the incessant growth of automotive and mobility services, accompanied by the escalating complexity of urban roadway environments, gives rise to a multitude of new challenges in terms of system communications within the IoV.  Various security issues that need to be addressed by the IoV framework are mentioned in  literature \cite{ref3}.  It is vital  to establish communication channels that are robust, secure, seamless, and scalable between vehicles and roadside units.
	
	One noteworthy concern arises from the haphazard and densely deployed cellular networks, as the received power from non-serving base stations can at times exceed that from the serving base stations (BSs) \cite{ref4}, \cite{ref_I_6}.  To address the intricate communication challenges, two fundamental aspects require attention: signal interference management and signal coverage enhancement. Signal interference management entails the implementation of diverse techniques, such as the utilization of higher-frequency wireless communication technologies \cite{ref_I_7}, \cite{ref5}, the adoption of adaptive modulation and coding techniques, and the incorporation of multiple antenna technologies ?\cite{ref_I_8}, \cite{ref6}. The primary advantage lies in the implementation of anti-jamming techniques, which can effectively mitigate or eliminate the adverse effects of signal interference on communications, thereby enhancing the overall quality of communication. However, it is important to acknowledge that signal interference management involves the utilization of intricate techniques and algorithms, requiring specialized expertise and skills for implementation, as well as ongoing system deployment, operation,  and maintenance. This may contribute to increased complexity in the deployment, operation, and maintenance of the system. Another aspect to consider is signal coverage enhancement, communication quality by incorporating wireless communication devices such as BSs and repeaters along roads to provide improved signal coverage and capacity.  The authors in \cite{ref7} presented a sophisticated proposition that advocates for a blockchain-based solution utilizing the Ethereum. This solution aims to establish resilient mechanisms for safeguarding communications in the IoV.  On a different note, the authors in \cite{ref8} devised a multi-level cluster-based satellite-terrestrial integrated communication model (MCSIC). Their objective is to achieve efficient real-time data communication in IoVs by leveraging data analytics. 
	
	However, many existing approaches primarily focus on communication between vehicles and fixed infrastructure, which poses challenges when it comes to establishing real-time connections with roadside units (RSUs) in the event of their destruction caused by natural disasters or human actions. Additionally, inadequate planning during the early stages of RSU construction can result in a plethora of obstacles that impede communication between vehicles and RSUs during subsequent development. These issues give rise to problems such as transmission delays, packet loss, and network congestion. Moreover, when a vehicle surpasses the communication range of the RSU, the ongoing service gets interrupted. Therefore, providing seamless connectivity between vehicles and RSUs in VANETs poses a formidable challenge.   
	
	In IoV networks,  both the UAVs and RSUs assume similar roles in terms of providing network coverage and facilitating vehicle connectivity. However, utilizing the UAVs for communication proves to be a significantly more effective option compared to the aforementioned methods, thanks to its inherent advantages such as robust visual range, extensive coverage, and adaptable deployment \cite{ref9}. The UAVs act as catalysts for amplifying the signal coverage of the system, enabling swift deployment into the network, and exhibiting minimal susceptibility to terrain limitations as they navigate through the skies with unrestricted mobility. By deploying at high altitudes, the UAVs can overcome the non-line-of-sight problem in wireless communication and monitor traffic flow, road congestion, and accidents from a heightened perspective, thereby providing timely real-time information about the traffic conditions to vehicles and drivers. 
	
	In this scenario, determining the optimal deployment locations for the UAVs has emerged as a new research focus. Existing studies often approach the UAV location problem from the perspectives of UAV resource management or vehicle communication quality requirements. However, in practical application environments, it is crucial to comprehensively consider both factors to enhance system performance and reliability.
	
	Expanding upon the aforementioned observations, this paper presents an advanced intelligent optimization algorithm, referred to as QoS-aware 3D coverage deployment of UAVs, aimed at improving vehicular networking with a specific emphasis on quality of service (QoS). This algorithm surpasses traditional genetic algorithms (GA) by integrating the K-means clustering algorithm and the K-means initialized grey wolf optimization (KIGWO) technique. The primary contribution of this study can be summarized as follows:
	
	\begin{itemize}
		\item   To create a realistic urban road environment, a 3D road scene is meticulously constructed. This scene comprises three distinct roads that are strategically positioned across various height layers. Furthermore, the vehicles within this simulation are confined to the road pavement, emulating real-world conditions accurately.
		
		\item  The K-means clustering algorithm is utilized to form cohesive groups by clustering the vehicles. The centroids of each cluster are then utilized to determine the horizontal position for initializing the UAVs. Furthermore, these centroids serve as the initial population for the genetic algorithm, allowing the optimization process to commence from a strategically determined starting point. 
		
		\item  The KIGWO technique is employed to determine the optimal fitness value of the UAV position by conducting an optimal search within the grey wolf population. This method involves randomly replacing one of the individuals in the initial population with the position of a grey wolf, thereby enhancing the algorithm's diversity and global search capability. 
		
		\item  The distances between the UAVs and vehicles are computed to determine whether the vehicles fall within the service range of the UAVs. Concurrently, the communication quality of the vehicles is assessed to verify if it meets the specified quality of service criteria. This evaluation encompasses examining whether the actual average uplink speed of the vehicles meets or exceeds the minimum predefined value and if the data rate satisfies the data capacity requirements of the UAVs. This assessment methodology guarantees the effectiveness of UAV deployment and ensures the desired communication QoS for the vehicles.
		
		\item  By adopting a competitive selection approach and incorporating crossover and mutation operations, we iteratively evaluate and select individuals based on their fitness values. Through multiple iterations, we identify the individual with the highest fitness value, resulting in the optimal solution that maximizes the total number of vehicles covered by the UAVs.
	\end{itemize}
	
	The remainder of this paper is organized as follows. Section II examines existing research on the topic of this thesis.  Section III encompasses the definition of the road scenario and system model, along with the formulation of the problem.  In Section IV,  we provide the appropriate algorithm to solve the above problem.  In Section V, simulations are performed and results are analyzed by using Matlab.  Finally, the paper is summarized in Section VI.   
	
	\section{Related Work}
	Existing research on UAV-assisted vehicular  networking systems (VNS) mainly focuses on two aspects: UAV resource management scheduling and optimization of quality of service perception.
	
	In the domain of UAV resource management scheduling, the primary focus lies in designing efficient UAV path planning algorithms that aim to maximize the communication links among vehicles. Additionally, research explores dynamic allocation of communication resources by UAVs to cater to the diverse communication requirements between different vehicles. In order to enhance network coverage in a collaborative manner, a scheme called Collaborative Network Coverage Enhancement (CONEC) was introduced in \cite{ref10}. This scheme incorporates various factors such as vehicle density, travel direction, and previous coverage information to determine the optimal deployment locations for UAVs. The placement of UAVs throughout the network is achieved by utilizing the particle swarm optimization (PSO) algorithm.  Meanwhile, another algorithm is further proposed to efficiently schedule UAVs for each time slot to minimize the wasted UAV service time.  Comparative analysis shows that the proposed CONEC scheme outperforms its counterparts by significantly improving the network performance of VANET in terms of metrics such as packet delivery ratio (PDR), hop counts (HOPs), end-to-end delay (EED), and throughput. In the literature \cite{ref11}, UAVs were employed for monitoring the routes of ground vehicles, and an integrated routing and surveillance (IRS) model was proposed. The objective of this model is to plan UAV paths in a way that minimizes both risk and travel cost. In \cite{ref12},  a deployment strategy for energy-efficient rechargeable UAVs under seamless coverage constraints was proposed. The strategy involves a two-stage joint optimization algorithm, which addresses both the optimal deployment of UAVs and the recharging and reconfiguration strategy for cyclic UAVs. The optimization process utilizes an efficient particle swarm optimization algorithm to achieve the desired results. In literature \cite{ref13}, in scenarios where the line of sight between UAVs and ground IoT devices could be obstructed by buildings or mountains, a joint multi-UAV path planning and transmission scheduling algorithm is devised. The primary objective of this algorithm is to maximize the number of computing tasks executed successfully and in a timely manner by UAVs while minimizing the overall energy consumption of the UAVs. Addressing the challenges of high deployment costs, significant communication transmission losses, and limited coverage of traditional fixed base stations in complex terrains, the concept of employing UAVs as airborne mobile edge computing (MEC) nodes was proposed in \cite{ref14}. To optimize the energy efficiency of the cluster computational offloading strategy while enabling a higher number of offloading services, a UAV swarm scheduling scheme is introduced. This scheme takes into consideration the total latency, energy consumption, path planning, and computation offloading. The objective is to achieve a UAV configuration that offers shorter path lengths, higher energy efficiency, and an increased capacity for providing offloading services. In \cite{ref15}, the authors presented a UAV-based online target detection system and devised an adaptive motion planner to facilitate autonomous and energy-efficient navigation. This innovative approach empowers UAVs to autonomously process data and execute suitable tasks within resource-constrained conditions. The developed system enables efficient utilization of resources while ensuring the accomplishment of tasks in an optimal manner. An intelligent moth flame optimization-based clustering algorithm (IMOC) was proposed in \cite{ref16} to address the topological constraints such as difficult routing, high mobility, node density, and frequent path failures in vehicular self-organizing networks. This technique provides maximum coverage to vehicle nodes with minimum resource consumption. The authors in \cite{ref17} developed  a method to plan UAV coverage in a two- dimensional region that improves previous methods, including restricting vehicle flights in the region of interest, requiring shorter computation time, which improves the coverage and avoids potential collisions. In addition to the above algorithms, the UAV trajectory problem can also be equated to a mechanical problem. In literature \cite{ref18}, the concept of artificial potential field (APF) is introduced to redefine the objectives in trajectory design. The UAV trajectory is represented as an ultra-thin, highly flexible density-variable rope that carries the UAV's velocity information. The original objective of optimizing system performance is transformed into minimizing the overall artificial potential energy along the rope. By applying principles of mechanics, the rope in the optimal solution is maintained in a balanced state, thereby achieving equivalent optimization of the UAV flight path.
	
	In parallel, researchers allocate communication resources of the UAVs based on the communication and QoS requirements among vehicles. This allocation process is executed with the goal of attaining optimal communication quality. The relationship between the air-ground path loss (PL) and the location of the UAV base station in the horizontal and vertical dimensions was modeled in \cite{ref19}, and the three-dimensional deployment problem is decoupled into two-dimensional horizontal placement and height determination connected through PL requirements and minimization, a genetic algorithm-based two-dimensional placement method was proposed to maximize the coverage of the user equipment (UEs) based on the user data rate requirements and the capacity constraints of the UAV base station to cover the total number of UEs. In reference \cite{ref20}, a randomly distributed vehicle user localization (RBL) method was presented, which utilizes the Received Signal Strength Indicator (RSSI) as a metric for vehicle user positioning. The objective is to minimize path loss and maximize ground vehicle coverage while fulfilling individual the user's QoS requirements and optimizing vertical positioning. In reference \cite{ref21}, the optimal positioning problem was divided into three parts: user mobility prediction, radio and QoS-aware user clustering, and UAV-BS positioning. A mechanism for the deployment of proactive radio QoS-aware UAV base stations was introduced. The positioning of UAV base stations in specific areas is determined based on the LTE network and QoS conditions. This approach aims to enhance the coverage range for mobile users and improve service quality. The authors in \cite{ref22} investigated  a UAV-assisted VANETs routing protocol that utilizes UAVs to enhance data routing and connectivity among ground vehicles. Simulation results demonstrated that hybrid communication between vehicles and UAVs is more suitable for VANETs compared to traditional vehicle-to-vehicle (V2V) communication. In literature \cite{ref23}, comprehensive research was conducted on UAVs serving as relay nodes for coverage extension, path loss compensation, and interference suppression. This study provided a valuable guidance for the design and analysis of UAV-assisted wireless communication systems. In reference \cite{ref24}, the authors simulated a scenario of a highway interchange bridge and propose an energy-aware UAV deployment considering the effects of bridge occlusion and vehicle movement in a three-dimensional setting. The authors in  \cite{ref25} introduced  a novel integrated air-ground network architecture that leverages dual-function UAVs to enhance communication and positioning performance in the network. The authors in \cite{ref26} presented  a UAV-assisted distributed routing framework (D-IoT) tailored for the Internet of Things (IoT) environment, focusing on providing the QoS.  The framework employed  a neural-fuzzy interference system for reliable and efficient path selection.
	
	The previous studies \cite{ref10}-\cite{ref26} primarily focus on UAV deployment locations either from the perspective of UAV resource management or communication QoS requirement. However, in practical application environments, it becomes crucial to comprehensively consider both aspects in order to enhance system performance and reliability. Consequently, this paper proposes a UAV-assisted intelligent optimization algorithm that incorporates quality of service awareness. By taking into account both UAV resource management and vehicle QoS requirements, the algorithm aims to optimize the deployment locations of UAVs, maximizing the number of vehicles covered by UAVs in the network while simultaneously fulfilling vehicle QoS requirements. The algorithm uses intelligent optimization techniques to adjust the deployment location of UAVs in combination with the location information of vehicles to maximize the performance of the vehicular networking system.
	
	\section{System Model and Problem  Formulation}
	In this section, we firstly describe the modeling assumptions for the scenario roads.  Then,  the air-to-ground (A2G) channel model between UAVs and vehicles and the UAV-to-vehicle coverage model are presented. Finally, we mathematically represent the problem based on these models. Table I lists some of the main notations used in the modeling of this system. Table I lists some of the main notations used in the modeling of this system.
	\begin{table}[hb]
		\centering
		\caption{Description of  Main  Symbols}
		\resizebox{0.5\textwidth}{!}{
			\begin{tabular}{|>{\centering\arraybackslash}m{2.5cm}|>{\centering\arraybackslash}m{6cm}|}
				\specialrule{0em}{0.5pt}{0.5pt}
				\hline
				Symbol & Definition \\
				\hline\hline
				$\mathcal{P}$ & Set of vehicles \\
				\hline
				$\mathcal{Q}$ & Set of UAVs \\
				\hline
				$L_{i}$ & Position of vehicle $i$ \\
				\hline
				$L_{j}$ & Position of UAV $j$ \\
				\hline
				$P_{i,j}^{LOS}$ & Probability of LOS link between vehicle $i$ and UAV $j$ \\
				\hline
				$L^{LOS}, L^{NLOS}$ & Path loss with LOS and NLOS links \\
				\hline
				$D_{ij}$ & Euclidean distance between vehicle $i$ and UAV $j$ \\
				\hline
				$L_{i,j}$ & Average path loss between vehicle $i$ and UAV $j$ \\
				\hline
				$P_{i,j}$ & Received power of vehicles in massive fading \\
				\hline
				$R_{up}$ & Average uplink rate of vehicles \\
				\hline
				$R_{min}$ & The minimum average ascending speed value of the vehicle conforming to QoS requirements \\
				\hline
			\end{tabular}
		}
	\end{table}
	
	\begin{figure}
		\begin{center}
			\includegraphics[width=3.4in,height=3.2in]{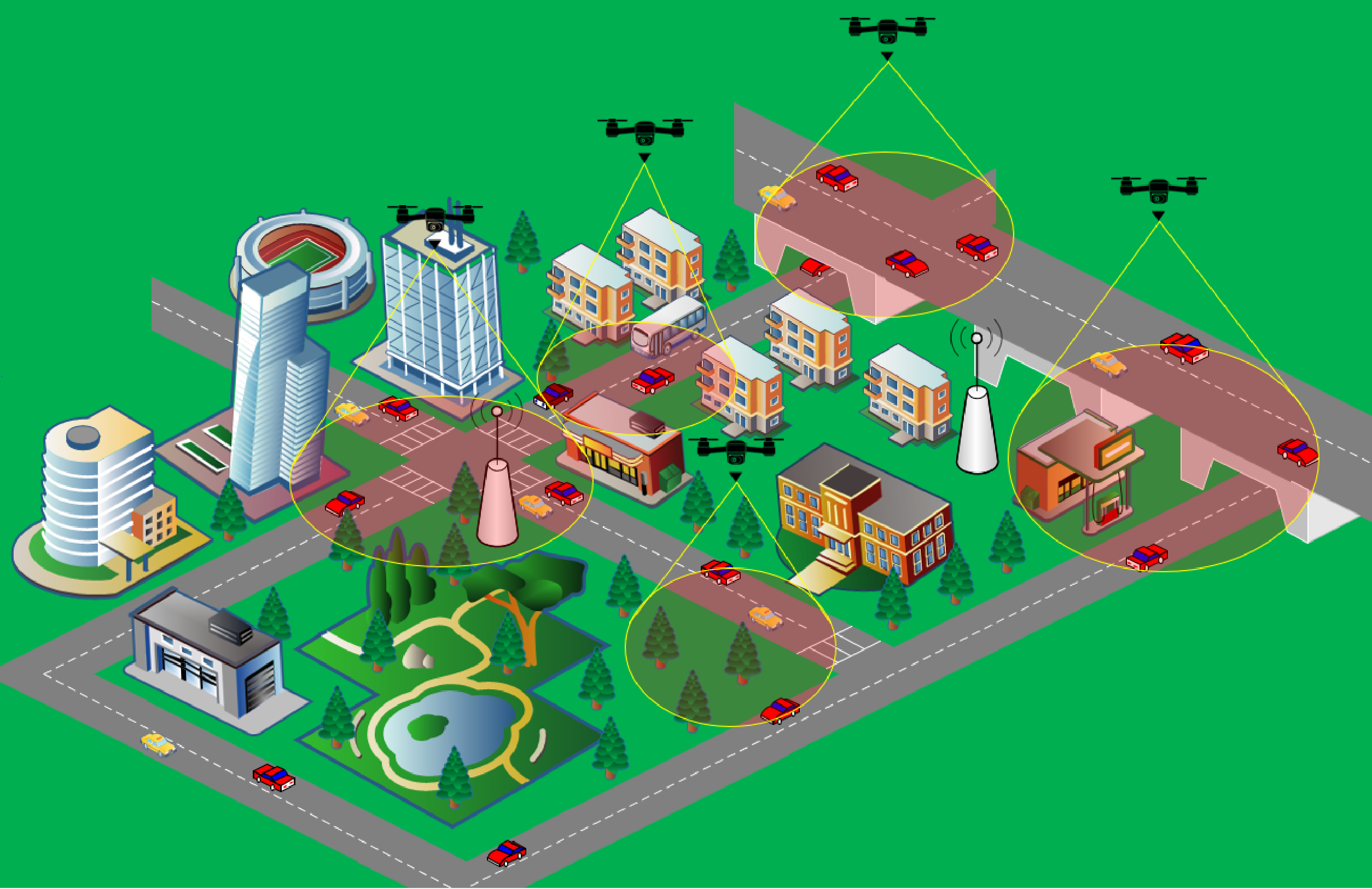}
			\label{fig1}\caption{System architecture of UAV-assisted IOV. }
		\end{center}
	\end{figure}
	As illustrated in Fig. 1,   the system realizes real-time data transmission and service support by intelligently connecting and cooperatively operating between the UAV  and vehicle user equipment (UE).  The UAV acts as a mobile node in the air to acquire information about the surrounding traffic environment and communicate with the UE. Similar to \cite{ref6},  our work does not include UAV energy consumption, we assume that the UAV has enough renewable energy to move from one place to another and the UAV is placed over all the obstacles.  The vehicles are equipped with onboard communication devices and sensing equipment, enabling them to gather information about their own status as well as data regarding surrounding vehicles and road conditions. Since the signal fading and path loss caused by obstacles such as buildings and interference from fixed ground communication base stations in complex urban environment \cite{ref27},  the signal degradation and distortion occur during transmission, resulting in reduced the  QoS and reliability.
	
	Let $\mathcal{{P}}$ denote the set of vehicles, and $i = 1,2,\ldots,\lvert \mathcal P \rvert$ be the number of each vehicle.  Let $\mathcal{{Q}}$ represent the set of UAVs, and let $j = 1,2,\ldots,\lvert \mathcal Q \rvert $ denote the individual identification number assigned to each UAV.  Taking the sea level (0 m) as the horizontal reference plane, the position of vehicle $i$ can be denoted as $\mathcal L_i = (x_i,y_i,h_i)$. The position of UAV $j$ can be represented by  $\mathcal L_j = (x_j,y_j,h_j)$. Since the vehicle is traveling on a road, its position $\mathcal L_i$ is limited to the road surface.
	
	\subsection{ Road Scene Model}
	The scenario simulates a viaduct in an urban roadway, set up with three roads, denoted as $Road_{m,m=1,2,3}$.The first road $Road_1$ is located above the other two roads and its bridge is shaded from the other two roads. Each road has a length of $l_{m,m=1,2,3}$ and a width of $w_{m,m=1,2,3}$. Each road in the system includes vertical slopes, indicating variations in height along the length of the road. The vertical altitude change of the road ranges from $0$ to $900$ meters.  It should be noticed that all the vehicles are randomly distributed on the roads.
	
	\subsection{ A2G Channel Model}
	A channel model is established for A2G communication scenarios in urban road environments, which takes into account the obstacles present in urban roads and the effects of multi-path propagation, and evaluates the performance of the communication system by means of metrics such as the average path loss and the line-of-sight (LoS) probability.
	\begin{figure}
		\begin{center}
			\includegraphics[width=3.4in,height=3.2in]{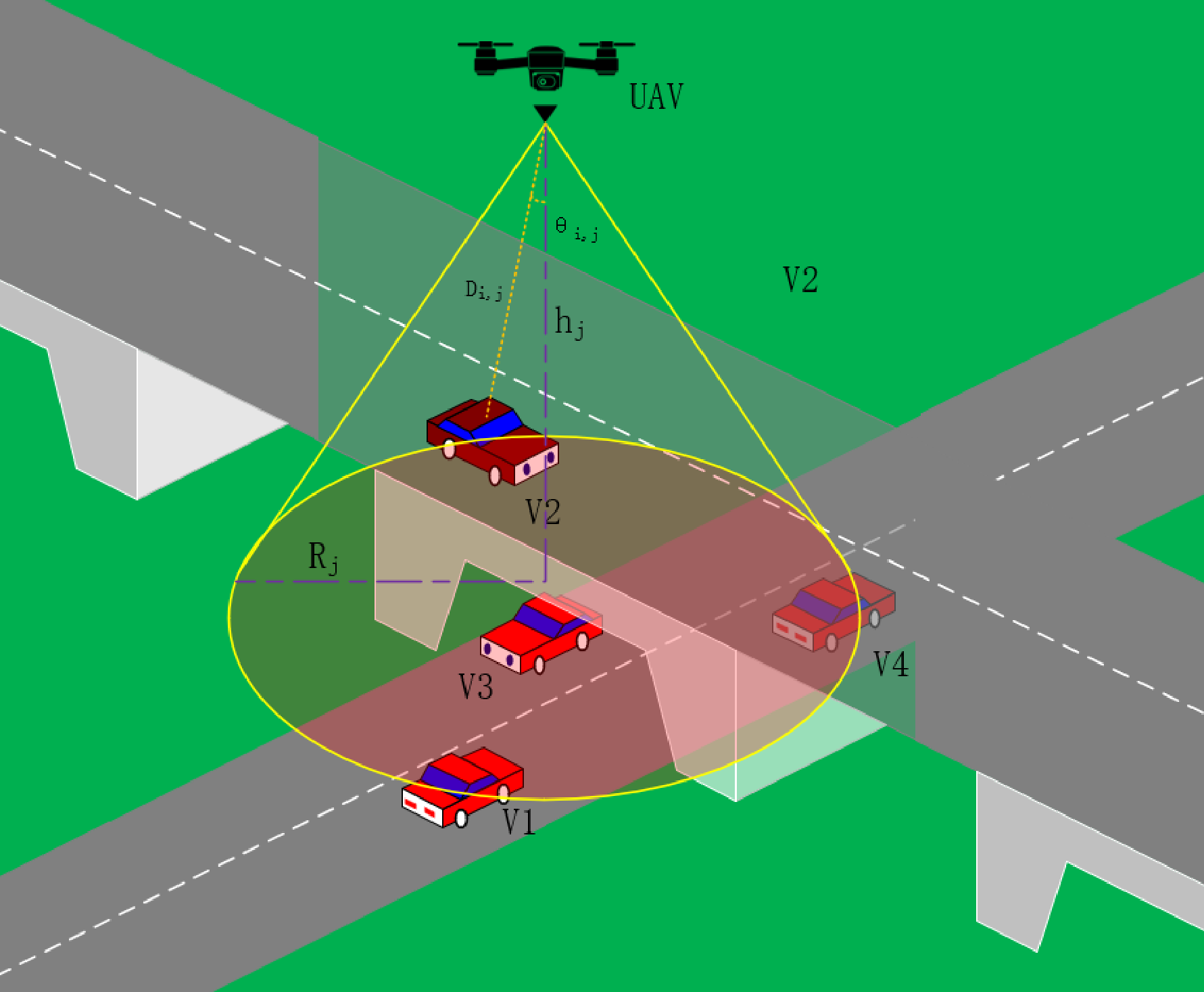}
			\label{fig2}\caption{Single UAV covers vehicle 3D scene .}
		\end{center}
	\end{figure}
	The LoS link probability of UAV-assisted IoV in urban road environments is the probability of the existence of a direct visible path between the UAV and the vehicle. The existence of LoS link means that there is no obstacle occlusion between the UAV and the vehicle, and better communication quality and performance can be realized. As shown in Fig. 2, the UAV covers the vehicle model in the scenario of road viaduct occlusion, where vehicle $V2$ is on the upper bridge deck.LOS communication is established between the UAV and vehicles $V1$ and $V2$, and Non-Line-of-sight (NLoS) communication is established with vehicles $V3$ and $V4$. In this scenario, although vehicle $V4$ is located in the coverage area of the UAV, the communication quality between vehicle $V4$ and the UAV is severely affected due to the occlusion of the bridge, and this interference leads to the fact that the average uplink  rate of vehicle $V4$ does not satisfy the requirements, thus preventing it from establishing a reliable communication connection with the UAV. We adopt an environment-based model to calculate the probability of a LOS link between vehicle $i$ and UAV $j$ using the following equation:
	\begin{align}
		\mathcal P_{i,j}^{LOS} = \frac{1}{1+\alpha e^{-\beta(90-\theta_{i,j}-\alpha)}}  \tag{1}
	\end{align}
	
	Where $(90-\theta_{i,j})$ is the elevation angle from vehicle $i$ to UAV $j$, which can be derived from the following equation:
	\begin{align}
		90-\theta_{i,j} = 90-arctan\left(\frac{h_j-h_i}{D_{i,j}}\right)*\frac{180}{\pi}  \tag{2}
	\end{align}
	
	$\alpha$ and $\beta$ are environment-dependent constant values indicating that the higher the number of buildings passing through, the lower the probability of LoS link. Communication loss refers to the signal loss or attenuation between UAVs and vehicles, and the magnitude of communication loss directly affects the communication quality and transmission performance between UAVs and vehicles. The model takes into account the distance, frequency, environmental conditions, and large-scale fading between UAVs and vehicles, then the path loss $L^{LOS}$ under $LOS$ link and the path loss $L^{NLOS}$ under the NLoS link can be given by 
	\begin{align}
		L^{LOS} = 20\log_{10}\left(\frac{4\pi f_c D_{i,j}}{c}\right) + \eta_{LOS} 
		\tag{3}  \\
		L^{NLOS} = 20\log_{10}\left(\frac{4\pi f_c D_{i,j}}{c}\right) + \eta_{NLOS} 
		\tag{4}
	\end{align}
	
	where $f_c$ is the carrier frequency, $c$ is the speed of light, and $\eta_{LOS}$ and $\eta_{NLOS}$ are the average additional losses that depend on the $LoS$ and $NLoS$ link environment, which takes into account the attenuation that a signal may encounter when propagating over a direct visible path. This attenuation may be due to absorption in the air, free space path loss, and antenna gain. The Euclidean distance between vehicle $i$ and UAV $j$ can be expressed as
	\begin{align}
		D_{i,j} = \sqrt{(x_i-x_j)^2+(y_i-y_j)^2+(h_i-h_j)^2}  \tag{5}
	\end{align}
	
	The average path loss between vehicle $i$ and UAV $j$ can be expressed as
	\begin{align}
		L_{i,j} = P_{i,j}^{LOS} L^{LOS}+\left(1-P_{i,j}^{LOS}\right) L^{NLOS}  \tag{6}
	\end{align}
	
	Large-scale fading is mainly considered in our setup model. Large-scale fading refers to the long-term fading effect in wireless communication systems due to path loss and environmental effects during signal propagation. The vehicle received power model in large-scale fading can be described by considering the communication gain $h_{i,j}$ and the transmit power $P_t$ between vehicle $i$ and UAV $j$:
	\begin{align}
		P_{i,j} = h_{i,j}*P_t \tag{7}
	\end{align}
	
	According to \cite{ref25} and \cite{ref_I_5},  the communication gain $h_{i,j}$ can be given by the following equation, where $h_0$ denotes the median average path gain at a reference distance of 1m.
	\begin{align}
		h_{i,j} = \frac{h_0}{D_{i,j}} \tag{8}
	\end{align}
	
	From the above analysis, we can calculate the average uplink rate of the vehicle $R_{up}$ , which needs to be greater than the minimum required value $R_{min}$ in the system model to meet the specific requirements.
	\begin{align}
		R_{up} = B_i \log_{2}\left(1+\frac{P_{i,j}}{\delta^2 \lambda L_{i,j}}\right)   \tag{9}
	\end{align}
	
	where $\delta^2$  is the Gaussian white noise power of UAV $j$ at the specified location, and the parameter $\lambda$ denotes the signal-to-noise ratio (SNR) gap between the actual modulation scheme and the theoretical Gaussian signal transmission, which is proposed to be fixed to a constant value using the QPSK modulation scheme in this model. $B_i$ is the allocated signal bandwidth obtained when all vehicles covered by UAV $j$ communicate with the UAV at the same time, which can be calculated by $B_i = B \phi\left(\lvert P \rvert \right)$, where $B$ is the total bandwidth and $\phi\left(\lvert P \rvert \right)$ is the channel utilization function \cite{ref28}.
	
	\begin{align}
		\phi\left(\lvert P \rvert \right) = a e^{-b\lvert P \rvert}  \tag{10}
	\end{align}
	
	$a$represents the initial channel utilization rate, set to $1$ (meaning that the available channel bandwidth is evenly distributed among all vehicles). Besides, $b$ represents the slope (indicating the rate at which the channel utilization rate decreases with each additional competing vehicle).  $\lvert P \rvert$ is the number of competing vehicles, and when $\lvert P \rvert$ is $0$, the channel utilization is $1$.
	\subsection{ UAV Coverage Model}
	A planar model of a UAV covering a vehicle is shown in Fig. 3.  The  coverage area of the UAV is conical in shape, the radius of the bottom circle is $R$ and the height is $H$. In this figure, vehicles $1$ and $2$ are located within the service area of the UAV, while vehicle $3$ is located outside the service area.
	
	\begin{figure}
		\begin{center}
			\includegraphics[width=3.4in,height=3in]{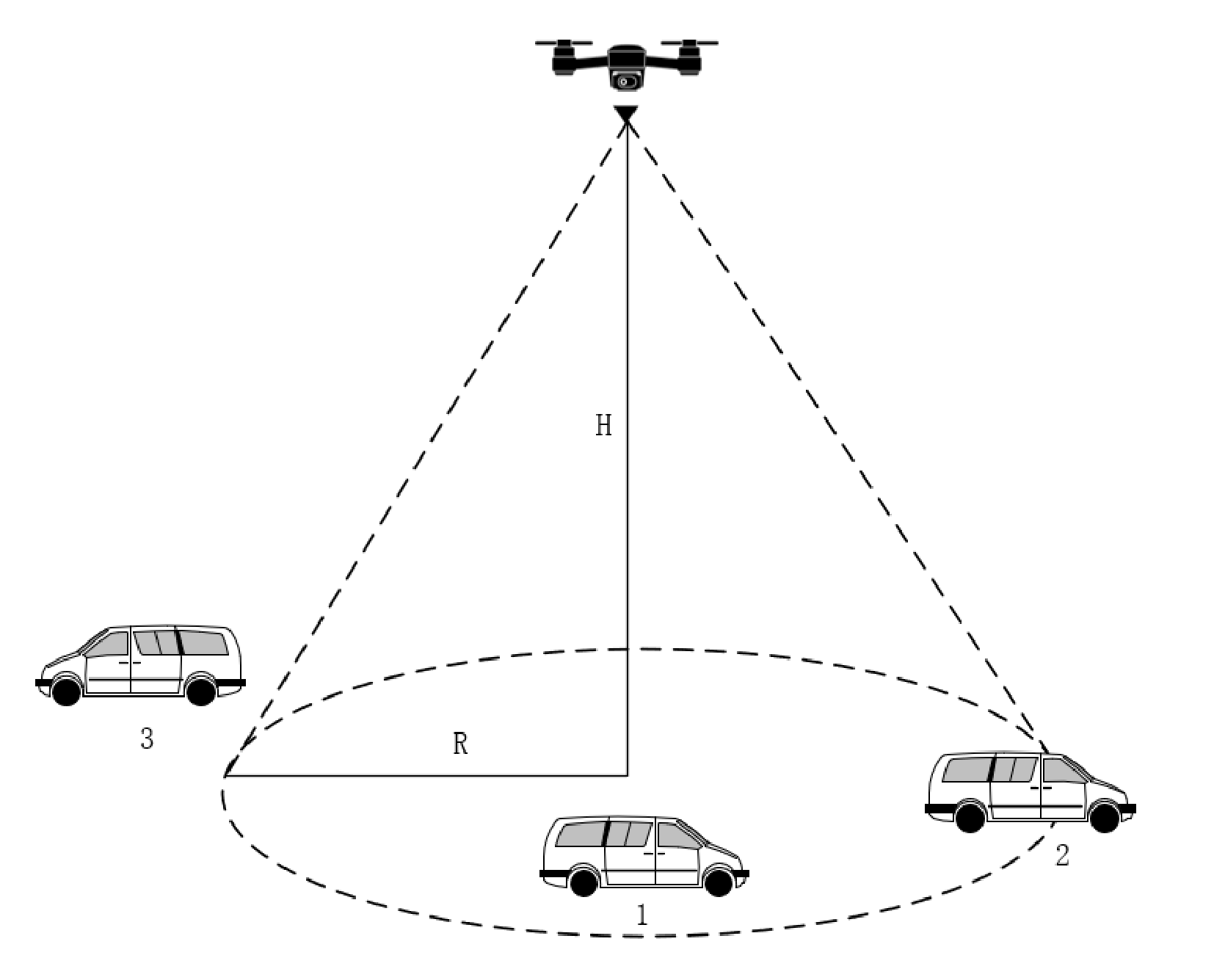}
			\label{fig3}\caption{UAV overlay vehicle plane model. }
		\end{center}
	\end{figure}
	
	Generally speaking,  when the UAV flies at a higher altitude, the field of view angle will become smaller and its maximum coverage radius will decrease accordingly.  In our model, to simplify the problem, we assume that the UAV's field of view angle $\theta$ is a fixed value of \SI{45}{\degree}, so the following relationship exists between the UAV's altitude and the maximum coverage radius
	\begin{align}
		R = H*\frac{H_{\alpha}}{\tan\frac{\theta}{2}},    \tag{11}
	\end{align}
	where $H_{\alpha}$ is a constant representing the linear relationship between the height of the UAV and the radius of coverage. The horizontal Euclidean distance between vehicle $i$ and UAV $j$ is $d_{i,j}=\sqrt{\left(x_i-x_j\right)^2 + \left(y_i-y_j\right)^2}$. A binary variable $\gamma_{i,j} \in \{0,1\}$ is used to indicate whether vehicle $i$ is covered by UAV $j$. If $d_{i,j} \leq R_j$,  then $\gamma_{i,j}$ takes the value of $1$ to indicate that the vehicle is within the coverage of the UAV and $0$ vice versa.
	
	Let $\{c_i\}_{i=1,2,\dots,\lvert P \rvert}$ denote the set of vehicle data rates and each vehicle has its corresponding data rate $c(i)$.The data rate of a vehicle is related to the vertical height of its location. In the simulated mountain road scenario, the vehicles are randomly distributed in three altitude layers, and the relationship between the vehicle data rate and vertical altitude is shown in Table II. $\{C_j\}_{j=1,2,\dots,\lvert Q \rvert}$ is the set of UAV data rate capacity, the sum of vehicle data rates covered by UAV $j$ cannot exceed the data rate capacity $C_j$ of this UAV.
	
	\begin{table}[hb]
		\centering
		\caption{Relation of Vehicle Height and Data Rate.}
		\resizebox{0.5\textwidth}{!}{
			\begin{tabular}{|>{\centering\arraybackslash}m{2.5cm}|>{\centering\arraybackslash}m{6cm}|}
				\specialrule{0em}{0.5pt}{0.5pt}
				\hline
				Height layer (m) & Data rate (MB) \\
				\hline
				0-300  & 0-2 \\
				\hline
				300-600& 2-3.5 \\
				\hline
				600-900& 3.5-5 \\
				\hline
			\end{tabular}
		}
	\end{table}
	
	\subsection{Problem Formulation}
	Based on the above scheme model,  we aim at maximizing the number of vehicles which are  covered under the premise of satisfying the UAV communication resource capacity and  communication QoS requirements. Thus, this optimization problem can be formulated as 
	\begin{alignat*}{2}
		\text{max} \quad & \sum_{i=1}^{\lvert P \rvert}\sum_{j=1}^{\lvert Q \rvert}\gamma_{i,j} \\
		\text{s.t.} \quad & C_1 : x_i \leq l_m , y_i \leq w_m                   \tag{12} \\
		& C_2 : R_{\text{up}}(i) \geq R_{\text{min}}                            \tag{13} \\
		& C_3 : \sqrt{\left(x_i-x_j\right)^2 + \left(y_i-y_j\right)^2} \leq R_j \tag{14} \\
		& C_4 : \sum_{i=1}^{\lvert P \rvert}c(i)*\gamma_{i,j} \leq C(j)          \tag{15} \\
		& C_5 : \sum_{j=1}^{\lvert Q \rvert}\gamma_{i,j} \leq 1                  \tag{16} \\
		& i\in \{1,2,\dots,\lvert P \rvert\},j \in \{1,2,\dots,\lvert Q \rvert\},m \in \{1,2,3\}
	\end{alignat*}
	
	where $C_1$ indicates that the vehicle location needs to be confined to the road pavement to ensure that the vehicle is within the legal travel area. $C_2$ indicates that the average uplink rate of the vehicle needs to be greater than the minimum required value to ensure the communication quality and system performance. $C_3$ indicates that the horizontal Euclidean distance between the vehicle and the UAVs needs to be within the coverage range of the UAVs. $C_4$ indicates that the system needs to comply with communication resource constraints, meaning that the total data transmission rate of vehicles covered by each UAV should not exceed the data capacity of the UAV. To avoid redundancy and duplicate coverage,  $C_5$ ensures that each vehicle is covered by one UAV.
	
	\section{Proposed QoS-Aware 3D UAV Coverage Deployment Algorithm}
	To address this optimization problem,   we propose an intelligent optimization algorithm for UAV-assisted IoV with the QoS awareness, referred to as Algorithm 1. The proposed algorithm improves on the traditional genetic algorithm (GA) by introducing the K-means clustering algorithm as the initialization method of the population in order to provide a better initial solution and increase the algorithm searching coverage of the space.  It is also combined with the improved KIGWO algorithm to increase the randomness.
	
	In the metaheuristic algorithm \cite{ref29}, the GA is a well-known algorithm. The fundamental components of GA include chromosome representation, fitness selection, and biological-inspired operators \cite{ref30}. The GA primarily consists of three operations: selection, crossover, and mutation.  In this algorithm we evaluate the degree of superiority of each individual by using the location of the UAV in space as the genetic coding and the number of vehicles covered by the UAV in the system as the fitness value. Based on the genetic coding of the UAV position, we employ the operations of selection, crossover and mutation in GA algorithm to manipulate the genes and thus generate a new UAV position solution.
	\begin{algorithm}
		\caption{QoS-Aware UAV-assisted  IoV Intelligent Optimization Algorithm (QoS-IOA)}
		\textbf{Input:} $l_{m,m=1,2,3}$, $w_{m,m=1,2,3}$, $K$, $D$, $\{P\}$, $\{Q\}$, $R_{\min}$, $p_c$, $p_m$ \\
		\textbf{Output:} $\{Best \_ {plan_j}\},R_j$ $j=1,2,\dots,\lvert Q \rvert$       \\
		\textbf{1:} Set up the road scene and obtain the vehicle location. The horizontal position information of the vehicle is stored in $\{u_i\}_{i=1,2,\dots,\lvert P \rvert}$, and the vertical height information of the vehicle is stored in $\{u_{h(i)}\}_{i=1,2,\dots,\lvert P \rvert}$. \\
		\textbf{2:}  Assign vehicle user data rates $\{c_i\}_{i=1,2,\dots,\lvert P \rvert}$ according to the altitude layer. Define the data capacity $\{C_j\}_{j=1,2,\dots,\lvert Q \rvert}$ of the UAV. \\
		\textbf{3:}  Create arrays $r$,$\hat{r}$ to store the number of vehicles covered by each UAV and the total number of vehicles covered by all UAV respectively (fitness value).\\
		\textbf{4:}  for z = 1:D\\
		\textbf{5:}  Use the K-means clustering algorithm (Algorithm 2) to obtain the initial population $\{population_1\}$.\\
		\textbf{6:}  Use the KIGWO algorithm (Algorithm 3) to randomly replace certain positions in $\{population_1\}$.\\
		\textbf{7:}  end \\
		\textbf{8:}  for $\hat{K}$ = $1:K$ \\
		\textbf{9:}  \quad for $z=1:D$ \\
		\textbf{10:}  \quad \quad for $i$=1:$\lvert P \rvert$ \\
		\textbf{11:}  \quad \quad Calculate the horizontal Euclidean distance $d_{i,j}$ and the coverage radius $R_j$ between the UAV and vehicle. \\
		\textbf{12:}  \quad \quad \quad if $d_{i,j} \leq R_j$ \\
		\textbf{13:}  \quad \quad \quad Assign the vehicle to the nearest drone and store the result in $\{O_j\}_{j=1,2,\dots,\lvert Q \rvert}$. \\
		\textbf{14:}  \quad\quad \quad end \\
		\textbf{15:}  \quad \quad end \\
		\textbf{16:}  \quad \quad for$j=1:\lvert Q \rvert$ \\
		\textbf{17:}  \quad \quad \quad if $\sum_{}^{}c(i)*\gamma_{i,j} \leq C(j)$ \& $R_{\text{up}}(i) \geq R_{\text{min}}$ \quad $i\in\{O_j\}$ \\
		\textbf{18:}  \quad \quad \quad $r(j)=length(\{O_j\})$ \\
		\textbf{19:}  \quad \quad \quad else \\
		\textbf{20:}  \quad \quad \quad $\hat{r}(z)  \leftarrow -100$ \\
		\textbf{21:}  \quad \quad \quad break; \\
		\textbf{22:}  \quad \quad \quad end \\
		\textbf{23:}  \quad \quad end \\
		\textbf{24:}  \quad Chromosome fitness $\hat{r}(z) = sum(r)$ \\
		\textbf{25:}  \quad end \\
		\textbf{26:}  Randomly generate competitors and select individuals with better fitness to store in $\{population_2\}$. \\
		\textbf{27:}  Randomly select two paternal chromosomes and replace genes according to the crossover rate $p_c$  to generate $\{population_3\}$. \\
		\textbf{28:}  Generate $\{population_4\}$ based on mutation rate $p_m$ \\
		\textbf{29:}  Repeat steps 9-25, and select the best fitness individual in the $\{population_4\}$ of this iteration and store it in $\{Alterbest\_plan\}$ \\
		\textbf{30:}  end \\
		\textbf{31:}  After K iterations, the individual with the highest fitness is selected in $\{Alterbest\_plan\}$ to obtain the optimal deployment position $\{Best \_ {plan_j}_{j=1,2,\dots,\lvert Q \rvert}\}$ and the corresponding coverage radius $R_j$ \\
		\textbf{32:} Return $\{Best \_ {plan_j}_{j=1,2,\dots,\lvert Q \rvert}\}$,  $R_j$ \\
	\end{algorithm}
	In Algorithm 1, we first input the predefined values of the corresponding variables, which include: the length and  the width of each road $l_{m,m=1,2,3}$, $w_{m,m=1,2,3}$. The number of loop iterations of the genetic algorithm $K$, the population size $D$, the set of vehicles $\{P\}$ and the set of UAVs $\{Q\}$, the minimum average uplink rate $R_{\min}$ that meets the requirements of the QOS,and $p_m$, $p_c$ which denote the mutation rate and the crossover rate, respectively. The road scene is built according to the provided road information,  and the vehicles are randomly distributed on the road surface, the horizontal position of the vehicles is stored in the collection $\{u_i\}_{i=1,2,\dots,\lvert P \rvert}$ and the vertical height information is stored in $\{u_{h(i)}\}_{i=1,2,\dots,\lvert P \rvert}$.  Then according to Table II,  determine the height layer where the vehicles are located and assign the vehicle user data rate $\{c_i\}_{i=1,2,\dots,\lvert P \rvert}$, define the data capacity of the UAV $\{C_j\}_{j=1,2,\dots,\lvert Q \rvert}$. In Step 3, two arrays $r$, $\hat{r}$ are created to store the number of vehicles covered by each UAV and the total number of vehicles covered by all UAVs (fitness score).
	
	For steps 4-7, the K-means clustering algorithm (Algorithm 2) is used to obtain the initial population $\{population_1\}$ and is combined with the KIGWO algorithm (Algorithm 3) to randomly replace certain positions in $\{population_1\}$.  Steps 8-30 show that the genetic algorithm undergoes K iterations, and in steps 10-15, we calculate the coverage radius of the UAVs and the horizontal Euclidean distance $d_{i,j}$  between the vehicle and each UAV according to Eq. (11),and if $d_{i,j}$ is less than or equal to the coverage radius of the UAVs, the vehicle is assigned to the UAV closest to it, and we set the assignment marker $\gamma_{i,j}$ or the current UAV to be 1, and store the result in $\{O_j\}_{j=1,2,\dots,\lvert Q \rvert}$. For steps 16-23, calculate whether the sum of the data rates of all the vehicles covered by each UAV exceeds the data capacity of that UAV, and whether the average uplink rate $R_{up}$ of the vehicles in the cluster is greater than $R_{min}$. Then update $r(j)$ to the number of vehicles covered by the UAV $j$ if the conditions are met. Instead assign a negative number to $\hat{r}$ and use break to jump out of the loop to return to step 4. Finally, add the elements of the  array $r$ and assign the result to $\hat{r}(z)$  for indicating the fitness of the chromosome.
	
	Through the selection operation, the adapted individuals are chosen from the population and copied to the next generation of the population. Then, the crossover operation is performed. According to the crossover rate $p_c$, a pair of paternal chromosomes is randomly selected to generate two zygotic chromosomes by exchanging gene fragments and replacing them into the next generation of the population. In addition, based on the mutation rate $p_m$, the chromosomes in the population are randomly mutated, the mutation sites are randomly selected and the mutation values are randomly generated. The last part of the algorithm is to recalculate the fitness of each individual in the population after mutation and select the individual with the highest fitness to be stored in $\{Alterbest\_plan\}$ as the suboptimal solution to the problem. When the $K$ iterations are over,  the individual with the highest fitness value among all the suboptimal solutions is selected as the optimal solution to the problem, and the algorithm returns the UAV deployment location $\{Best \_ {plan}\}$ that maximizes the number of covered vehicles in the system and the corresponding coverage radius $R$.
	
	\subsection{Cluster Vehicles into an Initial Population}
	The purpose of clustering is to divide a set of data patterns into disjoint clusters so that patterns in the same cluster are similar and patterns in different clusters are different \cite{ref31}.  The K-means clustering algorithm, on the other hand, is a commonly used unsupervised learning algorithm \cite{ref32}.
	
	In Algorithm 2, we input the vehicle horizontal position information $\{u_i\}_{i=1,2,\dots,\lvert P \rvert}$ as well as the number of vehicles $\lvert P \rvert$ and the number of UAVs $\lvert Q \rvert$. The parameters to be initialized include the maximum number of iterations Maxiter,  the precision $exp$ and the number of clusters $N$. The clustering centers are continuously updated by iteration until the clustering error is less than a given precision or the maximum number of iterations is reached.  In each iteration, the Euclidean distance between each vehicle and each cluster center is calculated and the vehicle is assigned to the cluster where the closest cluster center is located. Then, for each cluster, the average of all its data points assigned to that cluster is calculated and that average is used as the new cluster center.  Finally,  Algorithm 2 returns values as the location of the final cluster center $\{Z_j\}_{j=1,2,\dots,\lvert Q \rvert}$,and the index $\{cluster index\}$ of the vehicles contained in each cluster center.
	\begin{algorithm}
		\caption{Cluster Vehicles into an Initial Population}
		\textbf{Input:} $\{P\}$, $\{Q\}$, $\{u_i\}_{i=1,2,\dots,\lvert P \rvert}$ \\
		\textbf{Output:} $\{Z_j\}_{j=1,2,\dots,\lvert Q \rvert}$,$\{cluster index\}$      \\
		\textbf{1:}  Initialization parameters: Maxiter, accuracy $exp$, and number of clusters $N$. \\
		\textbf{2:}  Initialize cluster center $Z$. \\
		\textbf{3:}  for $iter = 1:M $\\
		\textbf{4:}  \quad Create an array of cells $\{cluster index\}$ to store the clustering results.\\
		\textbf{5:}  \quad for $i=1:\lvert P \rvert$\\
		\textbf{6:}  \quad \quad for $j=1:N$ \\
		\textbf{7:}  \quad \quad \quad  Calculate the distance of each vehicle to the cluster center and the vehicle is assigned to the cluster to which the nearest cluster center belongs. Store the vehicle number in the $\{cluster index\}$. \\
		\textbf{8:}  \quad \quad end \\
		\textbf{9:}  \quad end \\
		\textbf{10:}  \quad Update cluster center $Z$. \\
		\textbf{11:}  \quad for $j=1:N$ \\
		\textbf{12:}  \quad \quad $D$ stores the Euclidean distance between the location of each car and the center of each cluster in $Z$. \\
		\textbf{13:}  \quad end \\
		\textbf{14:}  \quad if $max(max(D) < exp$ \\
		\textbf{15:}  \quad \quad break; \\
		\textbf{16:}  \quad end \\
		\textbf{17:}  end \\
		\textbf{18:}  Return $\{Z_j\}_{j=1,2,\dots,\lvert Q \rvert}$,$\{cluster index\}$ \\
	\end{algorithm}
	\subsection{Randomly Replace Certain Positions}
	The grey wolf optimization (GWO) algorithm is a nature-inspired algorithm based on the hunting behavior and hierarchical leadership of grey wolves \cite{ref33}.  It seeks to find the optimal solution by simulating the cooperation and position updates among individual grey wolves. In nature, individuals in a gray wolf pack are divided into different hierarchies. $\alpha$ wolves are the leaders of the whole pack with the highest dominance and fitness. $\beta$ and $\delta$ wolves are the secondary leaders with slightly lower fitness than $\alpha$ wolves. $\omega$ wolves are the remaining individuals with relatively low fitness. Among them, $\alpha$ wolves represent the global optimal solution, $\beta$ and $\delta$ wolves represent the local optimal solution, and $\omega$ wolf packs represent other potential solutions. By constantly updating the position of the wolves, the algorithm searches for and obtains the solution with the largest fitness value.
	\begin{algorithm}
		\caption{ K-means Initialized Grey Wolf Optimization}
		\textbf{Input:} $\{P\}$, $\{Q\}$, $\{u_i\}_{i=1,2,\dots,\lvert P \rvert}$ \\
		\textbf{Output:} $\alpha$      \\
		\textbf{1:}  Initialization parameters: maximum number of iterations $T_{max}$,number of populations $W$. \\
		\textbf{2:}  Set the system boundaries and initialize the fitness values $\alpha_s = 0$,$\beta_s = 0$,$\delta_s = 0$ respectively.\\
		\textbf{3:}  Initialize the population position $\{initial_population\}$\\
		\textbf{4:}  for $\hat{t} =1:T_{max}$ \\
		\textbf{5:}  \quad for $i=1:W$ \\
		\textbf{6:}  \quad \quad Check the feasibility of individuals within the boundaries of the system. \\
		\textbf{7:}  \quad \quad The proximity between the UAV and the vehicle is used as the fitness measure to determine the individual fitness $fit(i)$. \\
		\textbf{8:}  \quad \quad if $fit(i)>\alpha_s$ \\
		\textbf{9:}  \quad \quad \quad  $\alpha_s = fit(i)$,\quad $\alpha = \{individual\}(i,:)$ \\
		\textbf{10:} \quad \quad if $fit(i)<\alpha_s \& fit(i)>\beta_s $ \\
		\textbf{11:} \quad \quad \quad  $\beta_s = fit(i)$,\quad $\beta = \{individual\}(i,:)$ \\
		\textbf{12:}  \quad \quad if $fit(i)<\alpha_s \& fit(i)<\beta_s \& fit(i)>\delta_s $ \\
		\textbf{13:}  \quad \quad \quad  $\delta_s = fit(i)$,\quad $\delta = \{individual\}(i,:)$ \\
		\textbf{14:}  \quad \quad end \\
		\textbf{15:}  \quad end \\
		\textbf{16:}  \quad Update the individual location \\
		\textbf{17:}  end \\
		\textbf{18:}  Return $\alpha$ \\
	\end{algorithm}
	
	In Algorithm 3,  we use the K-means clustering algorithm to initialize the population by using the initial horizontal position of the UAV as the center of each cluster,  selecting the first three individuals from the population as the initial positions of $\alpha$, $\beta$ and $\delta$ wolves, and the rest of the positions in the population are set to be the positions of $\omega$ wolves. Steps 4-16 are iterative updating process: for each individual, check whether it is beyond the environment boundary, and if so, restrict its position within the boundary. Calculate the fitness value based on the individual's position $fit(i)$, the fitness function is determined by calculating the distance between the vehicle position and the individual's position.  Where 8-13 denote updating the positions of $\alpha$, $\beta$ and $\delta$ wolves based on the individual fitness values, wherein the individual with the highest fitness is updated as an $\alpha$ wolf, the individual with the second highest fitness is updated as a $\beta$ wolf, and the individual with the third highest fitness is updated as a $\delta$ wolf. Step 15 updates the positions of the individuals according to the position update rules of the gray wolf optimization algorithm. Mathematically, the position update process is as follows:
	\begin{align}
		\vec{F} = \lvert\vec{C} \cdot \vec{Z}_p (T) - \vec{Z} (T) \rvert     \tag{17} \\
		\vec{Z} (T+1) = \vec{Z}_p (T) +\vec{E} \cdot \vec{F}       \tag{18}
	\end{align}
	
	where $\vec{Z}_p (T)$ and $\vec{Z} (T)$ are the position vectors representing the prey and gray wolf during the iteration process, respectively. $\vec{E}$ and $\vec{C}$ are the coefficient vectors controlling the magnitude of position update and the direction of position change, respectively. The estimation of $\vec{E}$ and $\vec{C}$ is as follows:
	\begin{align}
		\vec{E} &= 2 \vec{e} \cdot \vec{a}_1 - \vec{e} \tag{19} \\
		\vec{C} &= 2 \vec{a}_2 \tag{20}
	\end{align}
	
	where $\vec{a}_1$ and $\vec{a}_2$ take random vectors between 0 and 1, and $\vec{e}$ is a control vector whose nonlinear variation process can be described as:
	\begin{align}
		\vec{e}(T) = 2*\left[ 1- \left(\frac{T}{T_{max}}\right)^2\right]  \tag{21}
	\end{align}
	
	Finally,  the iterative update ends and returns the location of the optimal solution $\alpha$.
	\section{Simulation Results}
	\subsection{Simulation Setup}
	To evaluate the performance of the proposed algorithm, we conducted a series of simulation experiments. These experiments involved simulating vehicle communication and UAV resource allocation in a three-dimensional urban road scenario to validate the effectiveness and feasibility of the algorithm in optimizing UAV deployment locations. We designed multiple experimental scenarios that covered various vehicle densities and UAV resource distributions to comprehensively assess the algorithm's performance under different conditions.
	
	We initially set the road scenario range to $3km*3km*3km$. The Genetic Algorithm was iterated for 100 generations, with a population size of 10.The crossover rate $p_c$ and the mutation rate $p_m$ are $0.8$ and $0.1$, respectively. The minimum average uplink rate $R_{min}$ is set as $3.2bps$, and the transmission power $P_t$ is $50mv$ with a total bandwidth $B=3.6Mhz$. Carrier frequency $f_c=2Ghz$. Fig. 4 shows a schematic diagram of the road scenario, in which black lines are used to represent the road surface, while red circles represent the position of vehicles.
	\begin{figure}
		\begin{center}
			\includegraphics[width=3.4in,height=3.2in]{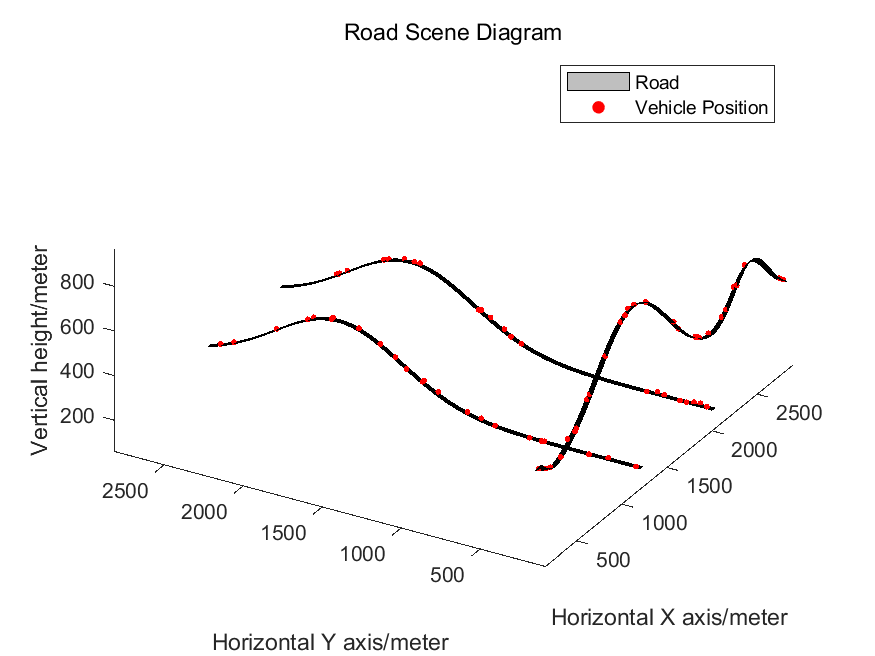}
			\label{fig4}\caption{The road scene diagram.}
		\end{center}
	\end{figure}
	\subsection{Simulation Results}
	\subsubsection{High-density Scenario}
	
	Due to the stochastic nature of the problem, we first set the number of UAVs in each scenario to 3 and perform 30 calculations. If we find an optimal solution that covers all the vehicles in the system, we will stop the computation. Otherwise, we increase the number of drones and continue to find an optimal solution.
	
	Fig. 5 shows the clustering results after K-means clustering algorithm when there are $80$ vehicles as well as $6$ UAVs in the system. In this figure, each circle represents a vehicle,  and different colored circles indicate different clusters. The number and distribution of the clusters reflect the spatial relationship and degree of aggregation between the vehicles, with larger clusters indicating the presence of a higher density of vehicles in the area, while smaller clusters indicate areas where the distribution of vehicles is sparser.
	\begin{figure}
		\begin{center}
			\includegraphics[width=3.4in,height=3.2in]{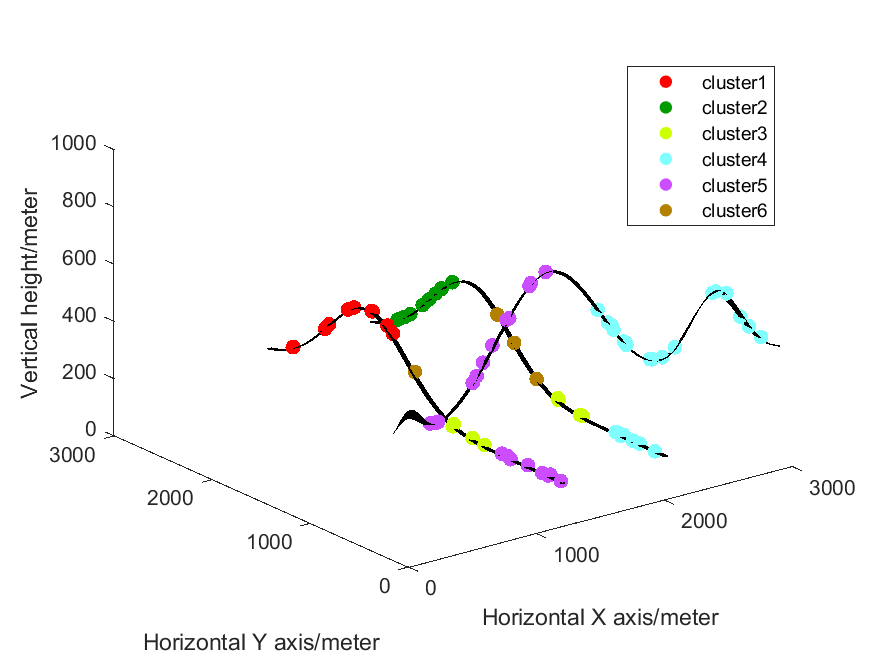}
			\label{fig5}\caption{Results of vehicle clustering in the high-density scenario.}
		\end{center}
	\end{figure}
	
	Fig. 6 shows the 3D effect of a UAV covering a vehicle in a high-density scenario, in which UAVs are represented by black triangles, and multiple UAVs can be seen distributed in the air at different heights and positions.  Each UAV forms a cone-shaped coverage with its location as the center. Through this 3D rendering, we can visualize multiple UAVs working in concert to cover the vehicle and provide services.
	\begin{figure}
		\begin{center}
			\includegraphics[width=3.4in,height=3.2in]{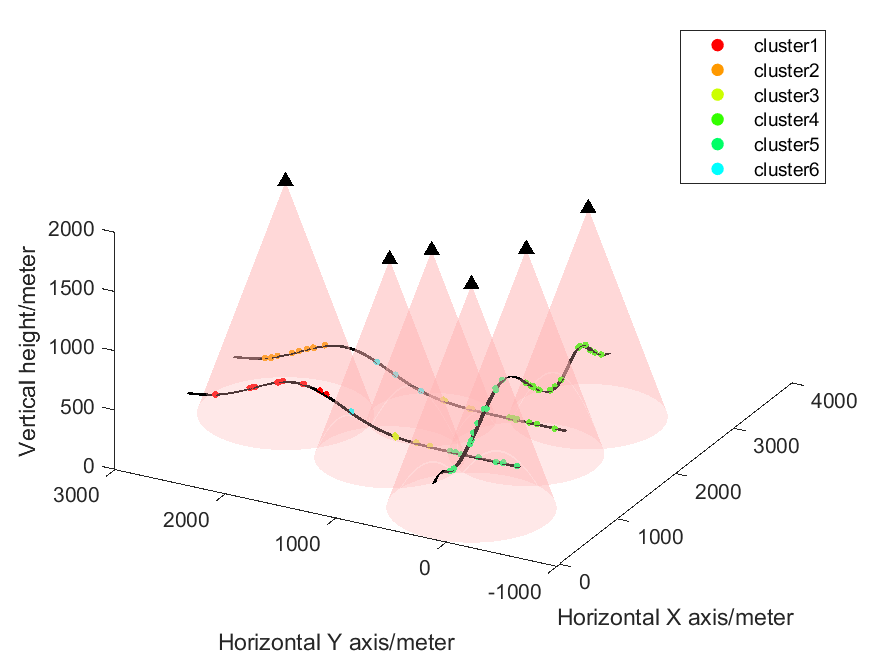}
			\label{fig6}\caption{The optimal 3D deployment of UAVs in the high-density scenario.}
		\end{center}
	\end{figure}
	
	The specific location information of the UAVs is shown in Table III, and it can be concluded from the analysis of the data in the table that the proposed UAV-assisted IoV intelligent optimization algorithm with the QoS awareness in this scenario enables the UAVs to obtain the optimal deployment location, which leads to $100\%$ coverage of the vehicles.
	\begin{table}[hb]
		\centering
		\caption{The  Optimal  Solution to  UAV 3D Deployment.}
		\resizebox{0.5\textwidth}{!}{
			\begin{tabular}{|>{\centering\arraybackslash}m{3cm}|>{\centering\arraybackslash}m{4cm}|>{\centering\arraybackslash}m{3cm}|}
				\specialrule{0em}{0.5pt}{0.5pt}
				\hline
				Horizontal position $(x_j,y_j)$ & Vertical height $h_j$ (m) & Cover radius $R_j$ (m) \\
				\hline
				(1775.00, 234.10)  & 1728.46 & 625.93 \\
				\hline
				(2015.00, 1216.00) & 1425.82 & 516.33 \\
				\hline
				(2650.50, 133.50)  & 1757.65 & 636.50 \\
				\hline
				(1562.83, 2301.13) & 1959.71 & 709.68 \\
				\hline
				(1202.43, 1169.50) & 1668.54 & 604.23 \\
				\hline
				(642.80, 135.00)   & 1883.32 & 682.01 \\
				\hline
			\end{tabular}
		}
	\end{table}
	
	\subsubsection{Low-density Scenario}
	
	In high-density scenarios, where there is a larger number of vehicles to be covered, the algorithm may have an easier time finding coverage solutions. However, in low-density scenarios, where the number of vehicles is smaller, the algorithm faces higher requirements for its effectiveness.
	
	Fig. 7 and Fig. 8  depict the clustering results of vehicles and the three-dimensional visualization of UAV coverage in a scenario with $30$ vehicles and $4$ UAVs. It can be clearly seen that in the low-density scenario, the distribution of vehicles is relatively sparse, and the vehicles are more inclined to form clusters with nearby vehicles. In this scene, our proposed algorithm can successfully achieve complete coverage of all vehicles in the scene. By simulating and analyzing the low-density scenario, we demonstrate the adaptability and effectiveness of the algorithm in different density scenarios, which shows that the algorithm can flexibly cope with different vehicle distributions and provide good coverage.
	
	\begin{figure}
		\begin{center}
			\includegraphics[width=3.4in,height=3.2in]{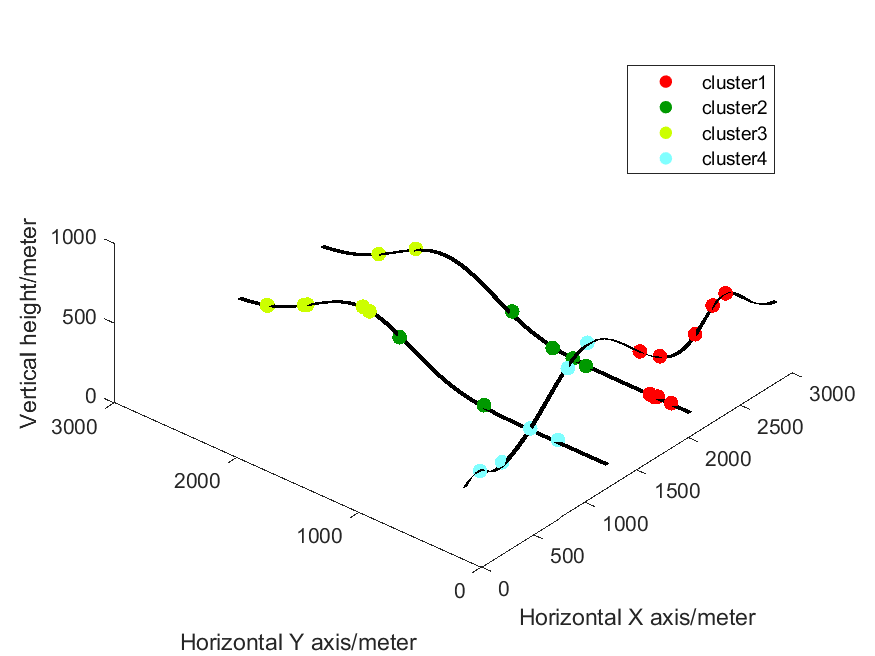}
			\label{fig7}\caption{Results of vehicle clustering in low-density scenario. }
		\end{center}
	\end{figure}
	
	\begin{figure}
		\begin{center}
			\includegraphics[width=3.4in,height=3.2in]{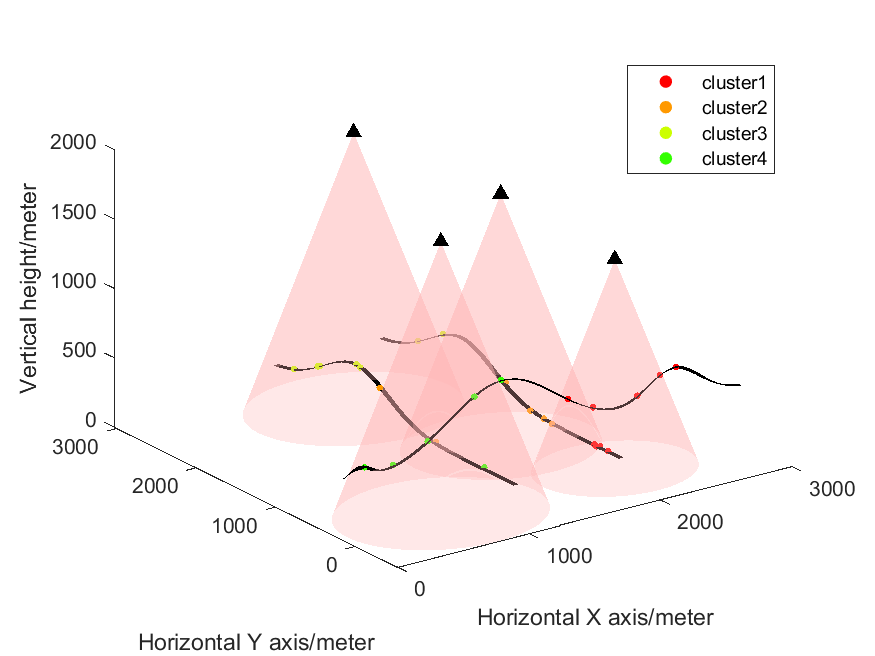}
			\label{fig8}\caption{The optimal 3D deployment of UAVs in low-density scenario. }
		\end{center}
	\end{figure}
	
	\subsubsection{Impact of Crossover and Mutation Rates on the UAV Coverage}
	
	The crossover rate $p_c$ and mutation rate $p_m$ are two important parameters in genetic algorithms. Fig. 9 shows the effects of different crossover rates and mutation rates on the system coverage rate with varying numbers of UAVs. It can be  observed from Fig. 9 that as the crossover rate decreases, the system coverage rate tends to increase. This is because a lower crossover rate reduces the exchange of ineffective genes, leading to a finer search range for the algorithm and potentially finding better solutions.  Conversely, as the mutation rate decreases, the system coverage rate decreases. This is because a lower mutation rate reduces the proportion of new genes introduced into the population, resulting in a smaller search range for the algorithm and potentially getting trapped in local optimal.
	
	\begin{figure}
		\begin{center}
			\includegraphics[width=3.4in,height=3.2in]{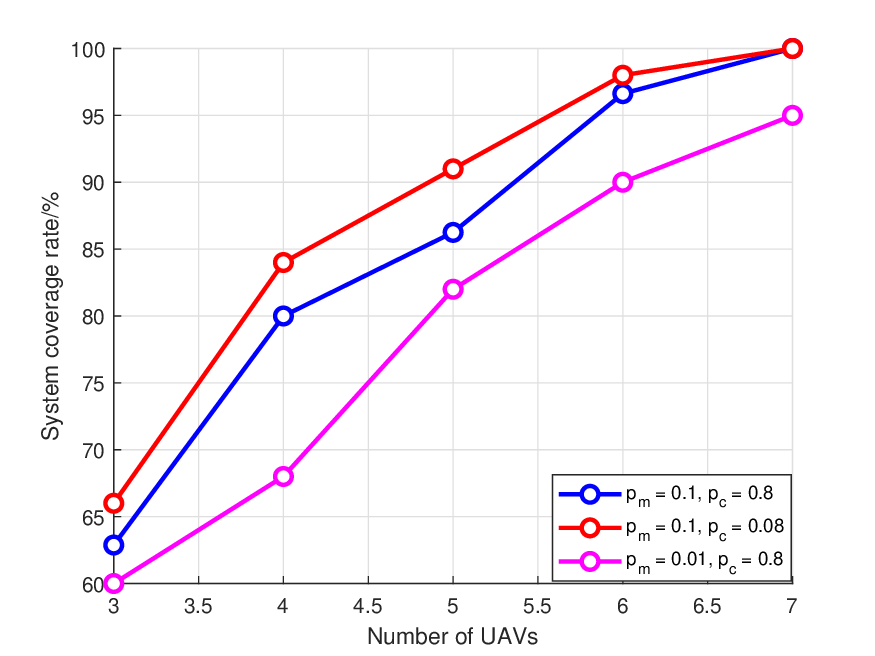}
			\label{fig9}\caption{Impact of the number of UAVs on the coverage under different $p_c$ and $p_m$. }
		\end{center}
	\end{figure}
	
	\subsection{Algorithm Performance Evaluation}
	
	We have taken the following model as a benchmark to evaluate the performance of the proposed algorithm, and the data in the figure are averaged over $30$ runs.
	
	\begin{itemize}
		\item GA \cite{ref30}: In the initialization process of traditional genetic algorithms, a certain number of chromosomes are randomly generated as the initial population. Each chromosome represents a candidate solution. Subsequently, selection, crossover, and mutation operations are performed on the population.
		\item PSO \cite{ref34}: Particle swarm optimization algorithm is a method of solving optimization problems by simulating the behavior of groups of organisms such as flocks of birds and schools of fish in nature. In particle swarm optimization algorithm, the position of each particle represents a solution in the solution space and the velocity of each particle represents its direction and speed in the search space. The algorithm searches for the optimal solution by constantly updating the position and velocity of the particles.
		\item SCA \cite{ref35}: The sine cosine algorithm (SCA) is a novel nature-inspired optimization algorithm proposed by Australian scholar Seyedali Mirjalili in 2016. This algorithm creates multiple random candidate solutions and utilizes the mathematical properties of sine and cosine functions. By adjusting their amplitudes, the algorithm balances its global exploration and local exploitation abilities during the search process, ultimately aiming to find the global optimal solution.
	\end{itemize}
	\subsubsection{The Number of UAVs}
	The number of UAVs is an important factor that influences the coverage rate.  Generally, it can be stated that a higher number of UAVs leads to a higher coverage rate. Fig. 10 shows the changes of the coverage rate of the four algorithms with different numbers of UAVs at $R_{min}=3.2bps$ in a high-density scenario.
	
	From the graph, it can be observed that under the same number of UAVs,  the proposed algorithm achieves the highest coverage rate. Furthermore, compared to other benchmarks,  this algorithm exhibits less randomness and provides a more uniform and accurate performance.
	\begin{figure}
		\begin{center}
			\includegraphics[width=3.4in,height=3.2in]{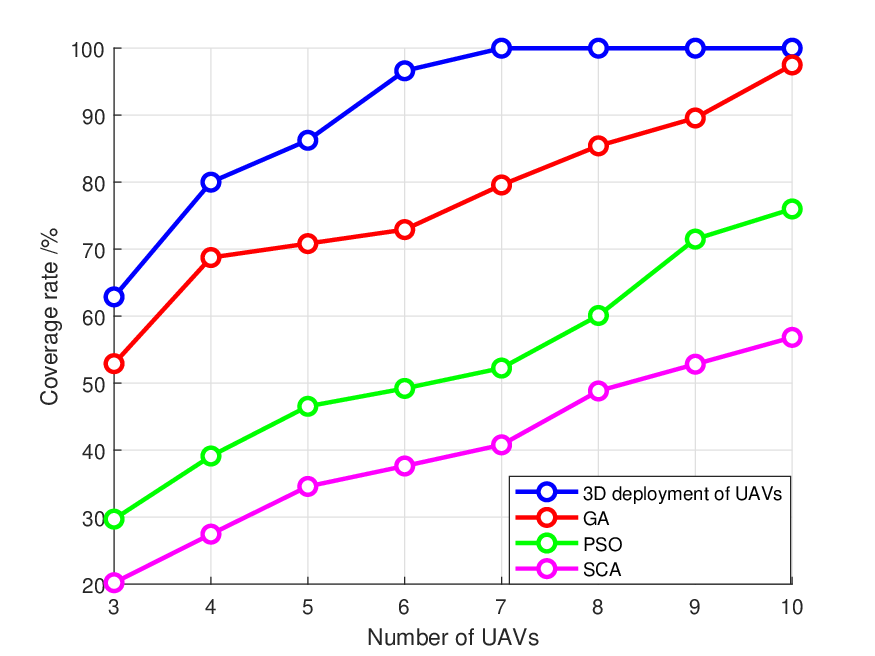}
			\label{fig10}\caption{The impact of the number of Uavs on coverage under different benchmarks.}
		\end{center}
	\end{figure}
	
	\subsubsection{Average Minimum Uplink Rate of Vehicles}
	
	Fig. 11 illustrates the impact of changes in the minimum average uplink speed of vehicles on the system coverage rate in a high-density scenario with $6$ UAVs covering vehicles. It can be observed from Fig. 11 that as the minimum average uplink speed increases, the system coverage rate decreases, and vice versa. This is because an increase in the minimum average uplink speed implies that each user is allocated more resources, resulting in a decrease in the number of users the system can serve, thus reducing the coverage rate. Conversely, a decrease in the minimum average uplink speed implies that each user is allocated fewer resources, leading to an increase in the number of users the system can serve and thus improving the coverage rate.
	
	\begin{figure}
		\begin{center}
			\includegraphics[width=3.4in,height=3.2in]{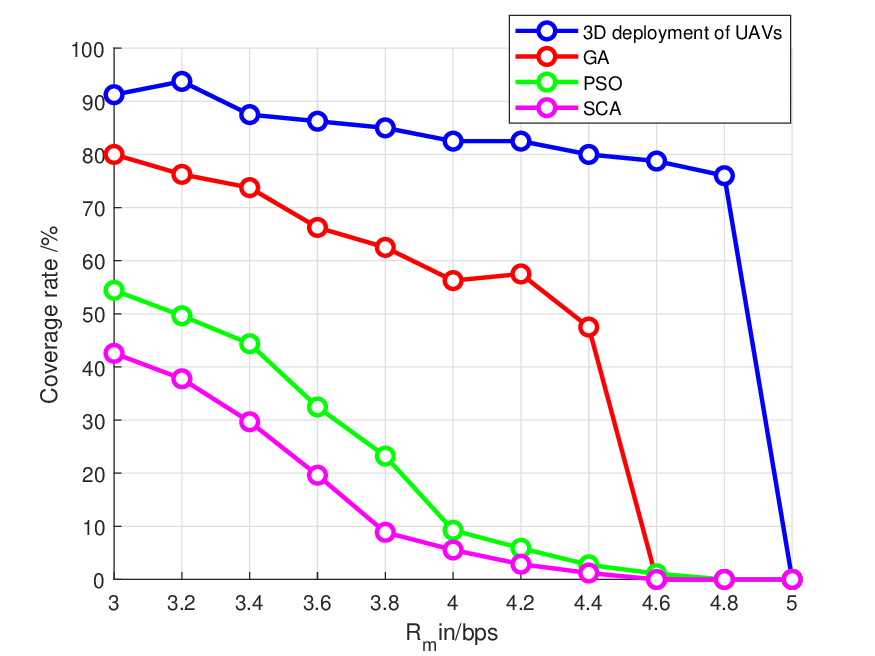}
			\label{fig11}\caption{The impact of $R_{min}$ on coverage under different benchmarks.}
		\end{center}
	\end{figure}
	
	Compared to the other benchmark algorithms,  the proposed algorithm  can provide a higher minimum average uplink speed for vehicles.  The communication is only disrupted when it exceeds a certain threshold. This indicates that the algorithm can establish more efficient communication with vehicles and provide better quality UAV services, thus better meeting the QoS requirements of vehicles.
	
	We have compared the performance of the proposed algorithm with GA, PSO and SCA algorithms. The results show that the proposed algorithm is more stable as well as adaptable compared to the benchmarks. This improvement allows the system to achieve higher vehicle coverage and improve the overall effectiveness and performance given the resource constraints .
	\section{Conclusion}
	
	In this work, we aim at  maximizing the number of vehicles which are  covered under the premise of satisfying the UAV communication resource capacity and  communication QoS requirements in IoV networks. To this end, we have  proposed  an intelligent 3D UAV coverage  algorithm in urban road environments,  which takes into account the communication resource capacity of the UAVs and the communication QoS requirements of the vehicles. By combining improved genetic algorithms with grey wolf optimization, the algorithm achieves the optimized deployment of UAV positions. In addition, abundant  experimental results have demonstrated that the proposed algorithm can effectively provide coverage to vehicles in IoV network while meeting their quality of service requirements.  Compared to other methods, this proposed algorithm  can reduce the number of required deployed UAVs while ensuring coverage for all vehicles. 
	
	In future research, the proposed algorithm can be further optimized and expanded to cope with increasingly complex urban environments and diverse traffic scenarios.  Firstly, for more complex urban environments, the algorithm can consider the impact of buildings, trees, and other obstacles on signal transmission. Secondly, considering the mobility of vehicles is crucial. Urban traffic is a dynamic system, and the algorithm design needs to take into account the prediction and trajectory planning of vehicles to adapt in real-time to changes in the positions and speeds of vehicles during UAV deployment and adjustment processes.

\end{document}